\documentclass[%
 reprint,
 amsmath,amssymb,
 aps,dvipdfmx
]{revtex4-1}
\usepackage{graphicx}
\usepackage{dcolumn}
\usepackage{bm}
\usepackage{color}

\renewcommand{\a}{\alpha}
\renewcommand{\b}{\beta}

\newcommand{\bea}{\begin{eqnarray}}
\newcommand{\eea}{\end{eqnarray}}
\newcommand{\f}[2]{\frac{#1}{#2}}
\newcommand{\eq}{&=&}
\newcommand{\nn}{\nonumber \\ }
\newcommand{\ve}{\varepsilon}
\renewcommand{\d}{\delta}
\newcommand{\area}{\int_{-\infty}^\infty }

\renewcommand{\l}{\lambda}
\newcommand{\p}{\partial}
\newcommand{\pp}[2]{\f{\p #1}{\p #2}}
\newcommand{\s}{\sigma}
\newcommand{\sref}[1]{Eq. (\ref{#1})}
\newcommand{\sd}{s}

\newcommand{\citeauthorname}[2]{{#1} {#2}}
\newcommand{\citebook}[4]{{#1} {\it #2} ({#3}, {#4}).}
\newcommand{\citepaper}[4]{{#1} {#3} ({#4}).}
\begin{document}

\preprint{APS/123-QED}

\title{Replica Analysis for the Duality of the Portfolio Optimization Problem
}

\author{Takashi Shinzato}
\email{takashi.shinzato@r.hit-u.ac.jp}
 \affiliation{
Mori Arinori Center for Higher Education and Global Mobility,
Hitotsubashi University, 
Tokyo, 1868601, Japan.}

\date{\today}

\begin{abstract}
In the present paper, 
the primal-dual problem consisting of the investment risk minimization problem 
and the expected return maximization problem 
in the mean-variance model 
is discussed using replica analysis. 
As a natural extension of the investment risk minimization problem under only a budget constraint that we analyzed in a previous study,  we herein consider a primal-dual problem in which the investment risk minimization problem with budget and expected return constraints is regarded as the primal problem, and the expected return maximization problem with budget and investment risk constraints is regarded as the dual problem. With respect to these optimal problems, 
we analyze a quenched disordered system involving both 
of these optimization problems 
using the approach developed in statistical mechanical 
 informatics, and confirm that both optimal portfolios can possess 
the primal-dual structure.
Finally, the results of 
numerical simulations are shown to {validate} 
the effectiveness of the proposed method.
\begin{description}
\item[PACS number(s)]
{89.65.Gh}, {89.90.+n}, {02.50.-r}
\end{description}
\end{abstract}
\maketitle

\section{Introduction}
The portfolio optimization problem, a mathematical finance problem, 
originating from the mean-variance model proposed by Markowitz in 1952 and 1959 \cite{Markowitz1952,Markowitz1959},
and the optimal strategy of diversification investment of the expected investment risk and the expectation of investment return, has been solved in several operations research studies
\cite{Bodie,Luenberger}. However, 
these previous studies considered only the analysis of annealed 
disordered systems in the framework of many-body complex systems. 
{The} analysis of a quenched disordered system {expected by the investor in practice}
has been investigated only rarely in operations research.
Recently, analysis of quenched disordered systems involving the portfolio 
optimization problem has been performed 
using replica analysis, the belief propagation method, and the asymptotic eigenvalue distribution of a random matrix, which were developed in several previous studies in 
interdisciplinary {research}  fields such as statistical mechanical informatics and 
econophysics \cite{Ciliberti1,Ciliberti2,Kondor,Pafka,Shinzato-SA2015,Shinzato-BP2015,VH,Shinzato-qw-fixed2016,Shinzato-ve-fixed2016}.
For example, 
Ciliberti et al. 
performed a replica analysis of the minimal investment risk for the absolute deviation model and 
the expected shortfall model for a quenched disordered system, and reported 
the phase diagram for the expected shortfall model in the zero temperature limit \cite{Ciliberti1,Ciliberti2}.
Kondor et al. evaluated the noise sensibility of the estimated error of 
the optimal solution with respect to several risk functions, such as variance, absolute 
deviation, expected shortfall, and maximum loss using numerical simulations 
\cite{Kondor}. Moreover, using numerical simulations, Pafka et al. showed that the optimal portfolio of each asset, which is the portfolio that can minimize the investment risk in the mean-variance model under only a budget constraint, has a normal distribution, and evaluated the ratio between the minimal expected investment risk and the in-sample risk \cite{Pafka}.
Shinzato reported that the minimal investment risk and its investment concentration in 
the mean-variance model under only a budget constraint are 
satisfied by the self-averaging property 
using a large-deviation approach and replica analysis,
and performed a quenched disordered analysis of this investment system \cite{Shinzato-SA2015}.
Furthermore, Shinzato et al. developed a fast algorithm for 
resolving the optimal solution that can
minimize each investment risk, including that in the mean-variance model, 
the absolute deviation model, and the expected shortfall model, 
using the belief propagation method \cite{Shinzato-BP2015}. 
Varga-Haszonits et al. determined the optimal portfolio 
that can minimize the variance between the whole return in each scenario and the expected return 
under budget and expected return constraints 
in a quenched disordered system using replica analysis \cite{VH}.
In addition, Shinzato performed an analysis of a quenched disordered system 
involving the investment risk minimization problem with budget and {investment concentration} 
constraints, and the {investment concentration maximization and/or minimization problem} with 
budget and investment risk using replica analysis, and clarified that the primal-dual structure also holds for a quenched system \cite{Shinzato-qw-fixed2016,Shinzato-ve-fixed2016}.

Quenched analysis has been examined extensively. In particular, the portfolio optimization problem under several constraints has been analyzed in a number of studies \cite{VH,Shinzato-qw-fixed2016,Shinzato-ve-fixed2016}.
However, the investment risk minimization problem with budget and expected return 
constraints, and the expected return maximization problem with budget and investment risk
constraints, have rarely been investigated.
Although Varga-Haszonits et al. 
solved the portfolio optimization problem under budget and expected 
return constraints 
using replica analysis \cite{VH}, 
the object function considered was 
defined as {half of}  the sum of the squares of the differences between the whole return in 
each scenario and the expected return, rather than using the variance of the whole return in each scenario.
In other words, 
since the object function considered was not 
the investment risk, 
the investment risk minimization problem with budget and expected return 
constraints, which is a natural extension of our previous study 
\cite{Shinzato-SA2015}, has not yet been examined. 
Furthermore, Shinzato first performed a replica analysis of the primal-dual problem 
with respect to the mean-variance model. 
From a unified viewpoint, it is also necessary to multidirectionally examine 
both portfolio optimization problems, i.e., the investment risk minimization problem with budget and expected return constraints and 
the expected return maximization problem with budget and investment risk
constraints.

The goal of the present paper is to perform a replica analysis of 
a primal-dual problem consisting of
the investment risk minimization problem and 
the expected return maximization problem in the mean-variance model, 
following the analytical approach 
used in previous studies 
\cite{Shinzato-qw-fixed2016,Shinzato-ve-fixed2016}. 
As a natural extension of our previous study \cite{Shinzato-SA2015}, which only analyzed an investment risk minimization problem under a budget constraint, we herein consider a primal-dual problem in which the investment risk minimization problem with budget and expected return constraints is regarded as the primal problem and the expected return maximization problem with budget and investment risk constraints is regarded as the dual problem. 
With respect to these portfolio optimization problems, 
using an analytical approach in statistical mechanical informatics, 
we analyze a quenched disordered system and determine whether both optimal solutions can possess the 
primal-dual structure.
Moreover, we compare with the results with 
those obtained through numerical simulations in order to validate the effectiveness of the proposed method.

The remainder of the present paper is organized as follows. In the next section, 
we describe the primal problem and the dual problem {handled herein}, where, for the sake of convenience, 
the primal problem is the investment risk minimization problem with budget and expected return constraints and the dual problem is the expected return maximization problem with 
budget and investment risk constraints. In Section 
\ref{sec3}, following the analytical procedure used in our previous study
\cite{Shinzato-SA2015},  
a replica analysis of a quenched disordered system involving the primal and dual problems is performed. Moreover, an annealed disordered system is analyzed using Lagrange's method of undetermined multipliers. 
In Section \ref{sec4}, in order to confirm the effectiveness of
the proposed method, we compare the results with those of numerical experiments. 
Finally, Section V presents conclusions and areas for future research.

\section{Model setting\label{sec2}}
First, we formulate the primal and dual problems in 
the mean-variance model {considered herein}.
\subsection{Primal problem\label{sec2.1}}
Similar to previous studies \cite{Ciliberti1,Ciliberti2,Kondor,Pafka,Shinzato-SA2015,Shinzato-BP2015,VH,Shinzato-qw-fixed2016},
the present paper considers 
a stable investment market that can handle the investment of $N$ assets.
Here, $w_i$ is a portfolio of assets $i(=1,2,\cdots,N)$, which can be described as $\vec{w}=(w_1,\cdots,w_N)^{\rm 
T}\in{\bf R}^N$, where  ${\rm T}$ represents 
the transpose of the matrix and/or vector.
Moreover, in order to simplify the discussion, 
we assume that short selling, i.e., 
$w_i\le0$ is allowed. Next, given $p$ scenarios (or periods), which are used in investing, 
the return rate of asset $i$ in scenario $\mu(=1,2,\cdots,p)$ can be described as
$\bar{x}_{i\mu}$, and the return of each asset is assumed to be 
independently distributed. Based on this assumption, we do not consider period correlation. Furthermore, the mean and variance of the return rate 
of asset $i$ are denoted as $r_i$ and $\sd^2$, respectively. Moreover, the mean $r_i$ is assumed to be a hyperparameter that is
independently and identically distributed with a probability 
distribution with a mean and variance of $m$ and $\s^2$, respectively. 
Thus, the investment risk ${\cal H}(\vec{w}|X)$ with respect to portfolio 
$\vec{w}$ is defined as follows:
\bea
\label{eq1}
{\cal H}(\vec{w}|X)\eq\f{1}{2N}\sum_{\mu=1}^p
\left(\sum_{i=1}^N\bar{x}_{i\mu}w_i-
\sum_{i=1}^Nr_iw_i
\right)^2\nn
\eq\f{1}{2}\sum_{\mu=1}^p
\left(\f{1}{\sqrt{N}}\sum_{i=1}^Nx_{i\mu}w_i\right)^2,
\eea
where the modified return rate $x_{i\mu}=\bar{x}_{i\mu}-r_i$ is used. 
Then, the mean and variance of the modified return rate are $0$ and 
$\sd^2$, respectively. Furthermore, 
the return rate matrix is given by $X=\left\{\f{x_{i\mu}}{\sqrt{N}}\right\}\in{\bf R}^{N\times p}$.
With respect to this object function, as 
budget and expected return constraints,
\bea
\label{eq2}N\eq\sum_{i=1}^Nw_i,\\
\label{eq3}NR\eq\sum_{i=1}^Nr_iw_i,
\eea
are used. The budget constraint in \sref{eq2} was used in previous studies 
\cite{Ciliberti1,Ciliberti2,Shinzato-SA2015,
Shinzato-BP2015,VH,Shinzato-qw-fixed2016}, and \sref{eq3} describes the expected return constraints. Moreover, 
$R$ is {a} coefficient characterizing the expected return. 
In addition, the feasible portfolio subset ${\cal P}(R)$ is given in terms of portfolio $\vec{w}$ and satisfies these two constraints:
\bea
{\cal P}(R)\eq
\left\{
\vec{w}\in{\bf R}^N\left|
N=\vec{e}^{\rm T}\vec{w},
NR=\vec{r}^{\rm T}\vec{w}
\right.
\right\},
\label{eq4}
\eea
where $\vec{e}=(1,\cdots,1)^{\rm T}\in{\bf R}^N$ is the constant vector 
and 
$\vec{r}=(r_1,r_2,\cdots,r_N)^{\rm T}\in{\bf R}^N$ is the mean vector.
Thus, 
the primal problem considered herein is the problem of determining the optimal portfolio $\vec{w}$
that can minimize the investment risk ${\cal H}(\vec{w}|X)$ in \sref{eq1}
in the feasible portfolio subset ${\cal P}(R)$ in 
\sref{eq4}. The minimal investment risk per asset $\ve$ is 
calculated as follows:
\bea
\ve\eq\mathop{\min}_{\vec{w}\in{\cal P}(R)}
\left\{\f{1}{N}{\cal H}(\vec{w}|X)\right\}.
\eea
Based on the argument in our previous study \cite{Shinzato-SA2015}, 
in the investment risk minimization problem imposing only a budget constraint,
the minimal investment risk per asset is $\ve=\f{\sd^2(\a-1)}{2}$, 
where the scenario ratio $\a=p/N$ is used. When 
the rank of the return rate matrix 
$X=\left\{\f{x_{i\mu}}{\sqrt{N}}\right\}\in{\bf R}^{N\times p}$, 
${\rm rank}(X)=N$, i.e., $p>N$, because 
the optimal solutions of the primal problem and the dual problem described later herein 
can be determined uniquely, 
and the argument is limited to $\a=p/N>1$.

Finally, there are two important considerations. First, 
using Lagrange's method of undetermined multipliers,
\bea
\ve\eq
\f{N}{2\vec{e}^{\rm T}J^{-1}\vec{e}}
\left\{1+
\f{\left(R-\f{\vec{r}^{\rm T}J^{-1}\vec{e}}{\vec{e}^{\rm T}J^{-1}\vec{e}}\right)^2}
{
\f{
\vec{r}^{\rm T}J^{-1}\vec{r}
}{\vec{e}^{\rm T}J^{-1}\vec{e}}
-
\left(\f{\vec{r}^{\rm T}J^{-1}\vec{e}}{\vec{e}^{\rm T}J^{-1}\vec{e}}\right)^2
}
\right\},
\label{eq6-1}
\eea
is solved analytically, where the Wishart matrix $J=XX^{\rm T}\in{\bf 
R}^{N\times N}$ is used. (See Appendix \ref{app-a} for details.) However, 
the computational complexity of the inverse of the Wishart matrix 
$J=XX^{\rm T}$ must be $O(N^3)$, 
and, in the analysis of a quenched disordered system, 
it is necessary to average
\sref{eq6-1} over the return rate matrix $X$ and the mean vector $\vec{r}$. Therefore,
it is not easy to directly evaluate the configuration average 
$E_{X,\vec{r}}[\ve]$, where $E_{X,\vec{r}}[f(X,\vec{r})]$ describes the 
expectation of $f(X,\vec{r})$ over the whole configuration of $X$ and $\vec{r}$.
As such, we will analyze the configuration average of the minimal investment risk per 
asset using replica analysis, which does not 
directly require evaluation of the inverse matrix in order to solve 
the configuration average.

Next, based on the definitions of the investment risk and expected return, 
we can separate the randomness in the expected return 
from the randomness in the investment risk.
In other words, in the mean-variance model, 
since the investment risk is defined as {half of}  the sum of the squares of 
the differences between 
whole return $\sum_{i=1}^N\bar{x}_{i\mu}w_i$ in scenario $\mu$ and 
expected return $\sum_{i=1}^Nr_iw_i$,
using the modified return rate $x_{i\mu}$,
we can remove the influence of the randomness of the expected return from the 
investment risk. Moreover, based on this formulation, 
the distributions of 
$x_{i\mu}$ and $r_i$ 
can be set independently.

\subsection{Dual problem}
In Subsection \ref{sec2.1},
similar to the primal problem, 
we formulate the 
dual problem as obtaining 
the optimal portfolio $\vec{w}$ that can maximize the expected return with imposed budget and investment risk constraints.
The expected return ${\cal H}'(\vec{w}|\vec{r})$ is defined as follows:
\bea
\label{eq6}
{\cal H}'(\vec{w}|\vec{r})\eq
\sum_{i=1}^Nr_iw_i,
\eea
where 
\bea
\label{eq7}
N\eq \sum_{i=1}^Nw_i,\\
\label{eq8}
N\ve'\eq\f{1}{2}\sum_{\mu=1}^p
\left(\f{1}{\sqrt{N}}\sum_{i=1}^Nw_ix_{i\mu}\right)^2,
\eea
are the budget and investment risk constraints. 
Equation (\ref{eq7}) is identical to the budget constraint given in \sref{eq2}, 
and \sref{eq8} describes the investment risk constraint. Moreover, 
$\ve'$ is a coefficient characterizing the investment 
risk, and the feasible portfolio subset ${\cal D}(\ve')$ is defined 
in terms of the portfolio 
$\vec{w}$, which satisfies {these two constraints:}
\bea
{\cal D}(\ve')\eq
\left\{
\vec{w}\in{\bf R}^N\left|
N=\vec{e}^{\rm T}\vec{w},
N\ve'=\f{1}{2}\vec{w}^{\rm T}J\vec{w}
\right.
\right\}.
\eea
Thus, the expected return per asset of the dual problem, $R'$, is 
calculated as follows:
\bea
R'\eq\mathop{\max}_{\vec{w}\in{\cal D}(\ve')}
\left\{\f{1}{N}{\cal H}'(\vec{w}|\vec{r})\right\}.
\eea
Moreover, using Lagrange's method of undetermined multipliers with 
respect to this dual problem,
\bea
\label{eq12}
R'\eq\sqrt{\f{\vec{r}^{\rm T}J^{-1}\vec{r}}
{\vec{e}^{\rm T}J^{-1}\vec{e}}
-
\left(\f{\vec{r}^{\rm T}J^{-1}\vec{e}}
{\vec{e}^{\rm T}J^{-1}\vec{e}}\right)^2
}
\sqrt{
\f{2\ve'\vec{e}^{\rm T}J^{-1}\vec{e}}{N}-1
}\nn
&&
+
\f{\vec{r}^{\rm T}J^{-1}\vec{e}}
{\vec{e}^{\rm T}J^{-1}\vec{e}},
\eea
is solved analytically. (See Appendix \ref{app-b} for details.) 
However, 
since it is {also} difficult to directly assess \sref{eq12} for a quenched 
disordered system, replica analysis is used.

Finally, there are two important considerations. First, the object function in \sref{eq1} in the primal problem corresponds to the second constraint in 
\sref{eq8} in the dual problem, and 
the object function in \sref{eq6} in the dual problem 
corresponds to the second constraint in \sref{eq3} in the primal problem. 
Second, the primal problem considered in the present paper is a natural extension of that in our previous study \cite{Shinzato-SA2015}. However, since we consider the expected return constraint in the present study, the proposed method is more practicable. 
The portfolio optimization problem with budget and expected return constraints 
has also been discussed \cite{VH}. However, since in that study,
the object function was not the investment risk, and the randomness of the asset returns in the investment risk and the expected return were different, 
it was not always a natural extension of our previous study \cite{Shinzato-SA2015}.  Furthermore, 
since the primal-dual structure is not handled analytically, 
it is more important to support the results of Varga-Haszonits et al. \cite{VH} by our findings in the present paper.

\section{Replica analysis\label{sec3}}
\subsection{Replica analysis for the primal problem}
In this section, we perform a replica analysis of a quenched disordered system 
involving the {primal} problem. Following our previous study 
\cite{Shinzato-SA2015},
the partition function for the canonical ensemble of this investment system 
of inverse temperature $\b$, $Z(R,X,\vec{r})$ is denoted as follows:
\bea
\label{eq11}
Z(R,X,\vec{r})
\eq\int_{\vec{w}\in{\cal P}(R)}d\vec{w}
e^{-\b{\cal H}(\vec{w}|X)},
\eea
where 
the investment risk 
${\cal H}(\vec{w}|X)$ in \sref{eq1} is regarded as the Hamiltonian, and the 
integral of 
$\vec{w}$ 
is regarded as the feasible portfolio subset ${\cal P}(R)$.

Thus, the minimal investment risk per asset $\ve$ is calculated 
from the following thermodynamic relation:
\bea
\ve\eq
\mathop{\min}_{\vec{w}\in{\cal P}(R)}
\left\{\f{1}{N}{\cal H}(\vec{w}|X)
\right\}\nn
\eq
-\lim_{\b\to\infty}\f{1}{N}\pp{}{\b}\log Z(R,X,\vec{r}).
\eea

The analysis of a quenched disordered system is performed as follows:
\bea
\phi(R)\eq\lim_{N\to\infty}\f{1}{N}
E_{X,\vec{r}}[\log Z(R,X,\vec{r})]\nn
\eq\mathop{\rm Extr}_\Theta
\left\{
\f{1}{2}(\chi_w+q_w)(\tilde{\chi}_w-\tilde{q}_w)+\f{q_w\tilde{q}_w}{2}\right.\nn
&&
-k-(R-m)\theta
+\f{\s^2\theta^2}{2}\chi_w
-\f{\a}{2}\log(1+\b\sd^2\chi_w)\nn
&&\left.-\f{\a\b\sd^2 q_w}{2(1+\b\sd^2\chi_w)}
-\f{1}{2}\log\tilde{\chi}_w+\f{\tilde{q}_w+k^2}{2\tilde{\chi}_w}
\right\},
\eea
where 
$\Theta=\left\{k,\theta,\chi_w,q_w,\tilde{\chi}_w,\tilde{q}_w\right\}$ 
is used. Moreover, the notation $\mathop{\rm Extr}_zf(z)$ describes 
the extremum of function $f(z)$ with respect to the variable $z$. (See 
Appendix \ref{app-c} for details.)

From these extremum conditions, 
as the results of the principal variables 
at inverse temperature $\b$, we obtain
\bea
\ve\eq\f{1}{2\b}+\f{\sd^2(\a-1)}{2}\left(1+\f{(R-m)^2}{\s^2}\right),\\
\label{eq15}
q_w\eq\f{\a}{\a-1}\left(1+\f{(R-m)^2}{\s^2}\right),\\
\chi_w\eq\f{1}{\b\sd^2(\a-1)}.
\eea
In the zero temperature limit, we obtain
\bea
\label{eq17}
\ve\eq\f{\sd^2(\a-1)}{2}\left(1+\f{(R-m)^2}{\s^2}\right),
\eea
where the investment risk per asset $\ve$ 
is a quadratic function of $R$. In addition, when $R=m$, the minimum value of 
$\ve$ is $\f{\sd^2(\a-1)}{2}$.
Now, if $R=m$ in \sref{eq3}, 
since the expected return constraint
is consistent with the budget constraint in \sref{eq2},
in practice, this result is consistent with 
the minimization of the investment risk with only a budget constraint being imposed.

Moreover, the 
Sharpe ratio, which measures the expected return with respect to the 
investment risk, i.e., $S=\f{R}{\sqrt{2\ve}}$, is derived as follows:
\bea
\label{eq20-1}
S\eq\f{1}{\sd\sqrt{\a-1}}\f{R}{\sqrt{1+\f{(R-m)^2}{\s^2}}}.
\eea
From $\pp{S}{R}=0$, $R=m+\f{\s^2}{m}$ and 
$\ve=\f{\sd^2(\a-1)}{2}\left(1+\f{\s^2}{m^2}\right)$ are calculated, 
and the maximum Sharpe ratio $S_{\max}$ is then obtained as follows:
\bea
S_{\max}\eq\f{\sqrt{m^2+\s^2}}{\sd\sqrt{\a-1}}.
\eea

Finally, we can also discuss the analysis of the annealed 
disordered system. We have 
\bea
\label{eq20}
\ve^{\rm OR}\eq
\f{\sd^2\a}{2}\left(1+\f{(R-m)^2}{\s^2}\right),\\
\label{eq21}
q_w^{\rm OR}\eq1+\f{(R-m)^2}{\s^2}.
\eea
(See Appendix \ref{app-e} for details.)
As the relationship between the minimal investment risk per asset $\ve$ in 
\sref{eq17} and the minimal expected investment risk per asset $\ve^{\rm OR}$ in \sref{eq20},
\bea
\ve&<&\ve^{\rm OR},
\eea
is obtained. Similarly, $S^{\rm OR}=\f{R}{\sqrt{2\ve^{\rm OR}}}<S$ also holds.

\subsection{Replica analysis for the dual problem}
In this subsection, we describe a replica analysis 
of a quenched disordered system involving a dual problem.
Following the above approach, the partition function for the canonical 
ensemble of this investment system with an inverse temperature $\b$, 
$Z(\ve',X,\vec{r})$, is defined as follows:
\bea
\label{eq23}
Z(\ve',X,\vec{r})
\eq\int_{\vec{w}\in{\cal D}(\ve')}d\vec{w}
e^{\b{\cal H}'(\vec{w}|\vec{r})},
\eea
where the expected return ${\cal H}'(\vec{w}|\vec{r})$ in \sref{eq6} is 
regarded as the Hamiltonian, and 
the integral of $\vec{w}$ is regarded as the feasible portfolio subset ${\cal D}(\ve')$.

Then, the maximum expected return per asset $R'$ is derived from 
the following thermodynamic relation:
\bea
\label{eq24}
R'\eq\mathop{\max}_{\vec{w}\in{\cal D}(\ve')}
\left\{\f{1}{N}{\cal H}'(\vec{w}|\vec{r})
\right\}\nn
\eq
\lim_{\b\to\infty}\f{1}{N}\pp{}{\b}\log Z(\ve',X,\vec{r}),
\eea
where 
in order to maximize the expected return ${\cal H}'(\vec{w}|\vec{r})$ in the dual problem, 
we do not use the description of the Boltzmann factor given in \sref{eq11}, but 
rather use that presented in \sref{eq23}. Note that we also use the thermodynamic relation 
given in \sref{eq24}.

Then, the analysis of the quenched disordered system is 
performed as follows:
\bea
\phi(\ve')\eq\lim_{N\to\infty}\f{1}{N}E_{X,\vec{r}}\left[
\log Z(\ve',X,\vec{r})
\right]\nn
\eq\mathop{\rm Extr}_\Theta
\left\{
\f{1}{2}(\chi_w+q_w)(\tilde{\chi}_w-\tilde{q}_w)+\f{q_w\tilde{q}_w}{2}
\right.\nn
&&-k+\theta\ve'+
\b m+\f{\s^2\b^2}{2}\chi_w
-\f{\a}{2}\log(1+\theta\sd^2\chi_w)
\nn
&&
\left.
-\f{\a\theta\sd^2q_w}{2(1+\theta\sd^2\chi_w)}
-\f{1}{2}\log\tilde{\chi}_w
+\f{\tilde{q}_w+k^2}{2\tilde{\chi}_w}
\right\}.
\eea
(See Appendix \ref{app-d} in for details.)

From the extremum conditions, as 
the results of the principal variables at inverse temperature $\b$, we have
\bea
R'\eq m+\s(\b\s\chi_w),\\
q_w\eq \f{\a}{\a-1}
\left(1+\b^2\s^2\chi_w^2\right),\\
\chi_w\eq\f{1}{\theta\sd^2(\a-1)}.
\eea
Furthermore, from 
$\ve'=
\f{\a\sd^2\chi_w}{2(1+\theta\sd^2\chi_w)}+
\f{\a\sd^2q_w}{2(1+\theta\sd^2\chi_w)^2}
=\f{1}{2\theta}+\f{\sd^2(\a-1)}{2}\left(1+\b^2\s^2\chi_w^2\right)$, 
$\b$ and $\theta$ are satisfied by the following relation:
\bea
\left(\f{\b\s}{\theta\sd^2(\a-1)}\right)^2\eq
\f{2}{\sd^2(\a-1)}\left(\ve'-\f{1}{2\theta}\right)-1.
\eea

In the zero temperature limit, 
since the right-hand side is $O(1)$, $\b/\theta\sim O(1)$ holds. Then, 
\bea
\label{eq30}
R'\eq m+\s\sqrt{\f{2\ve'}{\sd^2(\a-1)}-1},\\
\label{eq31}
q_w\eq \f{\a}{\a-1}\f{2\ve'}{\sd^2(\a-1)},
\eea
are obtained. Moreover, the Sharpe ratio $S=\f{R'}{\sqrt{2\ve'}}$ is solved 
as follows:
\bea
\label{eq34-1}
S\eq\f{m+\s\sqrt{\f{2\ve'}{\sd^2(\a-1)}-1}}{\sqrt{2\ve'}}.
\eea
In addition, from $\pp{S}{\ve'}=0$,
$\ve'=\f{\sd^2(\a-1)}{2}\left(1+\f{\s^2}{m^2}\right)$
 and $R'=m+\f{\s^2}{m}$ are calculated. Then,  
the maximum Sharpe ratio $S_{\max}$ is given as 
\bea
\label{eq34}
S_{\max}\eq\f{\sqrt{m^2+\s^2}}{\sd\sqrt{\a-1}}.
\eea

Finally, three points should be noted here. First,
the previous subsection and this subsection describe 
the primal-dual structure. When we derive $R$ from \sref{eq17}, 
and set $R=R'$ and 
$\ve=\ve'$, 
\sref{eq30} is obtained. Similarly, $R=R'$ and 
$\ve=\ve'$ are set, and $q_w$ in \sref{eq15} is consistent with 
$q_w$ in \sref{eq31}. In other words, for a quenched disordered system,  
the optimal portfolio that can minimize the investment risk under a fixed expected 
return is consistent with the optimal portfolio that can maximize the expected return under a fixed
investment risk.

Next, the optimal portfolio that can minimize the expected return 
under a fixed investment risk is also solved using replica analysis. 
Therefore, the minimum expected return per asset $R''$ is
\bea
R''\eq\mathop{\min}_{\vec{w}\in{\cal D}(\ve')}
\left\{
\f{1}{N}{\cal H}'(\vec{w}|\vec{r})
\right\}\nn
\eq\lim_{\b\to-\infty}\f{1}{N}
\pp{}{\b}\log Z(\ve',X,\vec{r})\nn
\eq m-\s\sqrt{\f{2\ve'}{\sd^2(\a-1)}-1}.
\label{eq36}
\eea
In other words, there exists a portfolio for which the expected return is not less than \sref{eq36}.

Finally, we discuss the analysis of an annealed disordered system:
\bea
\label{eq34}
R'^{\rm OR}
\eq m+\s\sqrt{\f{2\ve'}{\sd^2\a}-1},\\
\label{eq35}
q_w^{\rm OR}\eq\f{2\ve'}{\sd^2\a}.
\eea
(See Appendix \ref{app-f} for details.) From Eqs. (\ref{eq34}) 
and (\ref{eq35}), we obtain
\bea
q_w^{\rm OR}\eq
1+\f{(R'-m)^2}{\s^2},
\eea
which is consistent with the finding in \sref{eq21}. In addition, from Eqs. (\ref{eq30}) and 
(\ref{eq34}), $R'>R'^{\rm OR}$ holds, and 
Sharpe ratios $S=\f{R'}{\sqrt{2\ve'}}$ and 
$S^{\rm OR}=\f{R'^{\rm OR}}{\sqrt{2\ve'}}$ are confirmed to be satisfied for the case in which 
$S>S^{\rm OR}$.

\section{Numerical experiments\label{sec4}}
In this section, 
we investigate the validity of the proposed method through numerical experiments.
The Wishart matrix $J=XX^{\rm T}\in{\bf R}^{N\times N}$ can be defined in terms of the return rate matrix $X$ and the mean vector $\vec{r}$, as follows: 
\bea
\ve(R,X,\vec{r})\eq
\f{N}{2\vec{e}^{\rm T}J^{-1}\vec{e}}
\left\{1+
\f{\left(R-\f{\vec{r}^{\rm T}J^{-1}\vec{e}}{\vec{e}^{\rm T}J^{-1}\vec{e}}\right)^2}
{
\f{
\vec{r}^{\rm T}J^{-1}\vec{r}
}{\vec{e}^{\rm T}J^{-1}\vec{e}}
-
\left(\f{\vec{r}^{\rm T}J^{-1}\vec{e}}{\vec{e}^{\rm T}J^{-1}\vec{e}}\right)^2
}
\right\},\qquad\\
R'(\ve',X,\vec{r})\eq\sqrt{\f{\vec{r}^{\rm T}J^{-1}\vec{r}}
{\vec{e}^{\rm T}J^{-1}\vec{e}}
-
\left(\f{\vec{r}^{\rm T}J^{-1}\vec{e}}
{\vec{e}^{\rm T}J^{-1}\vec{e}}\right)^2
}
\sqrt{
\f{2\ve'\vec{e}^{\rm T}J^{-1}\vec{e}}{N}-1
}\nn
&&
+
\f{\vec{r}^{\rm T}J^{-1}\vec{e}}
{\vec{e}^{\rm T}J^{-1}\vec{e}}.
\eea
Based on this, the $C$ return rate matrices, $X^1,X^2,\cdots,X^C\in{\bf 
R}^{N\times p}$ and the $C$ mean vectors, $\vec{r}^1,\vec{r}^2,\cdots,\vec{r}^C\in{\bf R}^N$,
\bea
\ve\eq\f{1}{C}\sum_{c=1}^C\ve(R,X^c,\vec{r}^c),\\
R'\eq\f{1}{C}\sum_{c=1}^CR'(\ve',X^c,\vec{r}^c),
\eea
are estimated, where the elements of the $c$th return rate matrix 
$X^c=\left\{\f{x_{i\mu}^c}{\sqrt{N}}\right\}\in{\bf R}^{N\times p}$, 
$x_{i\mu}^c$, have independent and identical probability distributions having a mean of 0 and a variance of $\sd^2$, 
and the component of the 
$c$th mean vector $\vec{r}^c=(r_1^c,\cdots,r_N^c)^{\rm T}\in{\bf R}^N$, 
$r_{i}^c$ has an independent and identical probability distribution having a mean of $m$ and a variance of $\s^2$.
Moreover, the investment concentration $q_w$ and the Sharpe ratio $S$ are also estimated.

\begin{figure}[tbh]
\begin{center}
\includegraphics[width=0.9\hsize,angle=0]{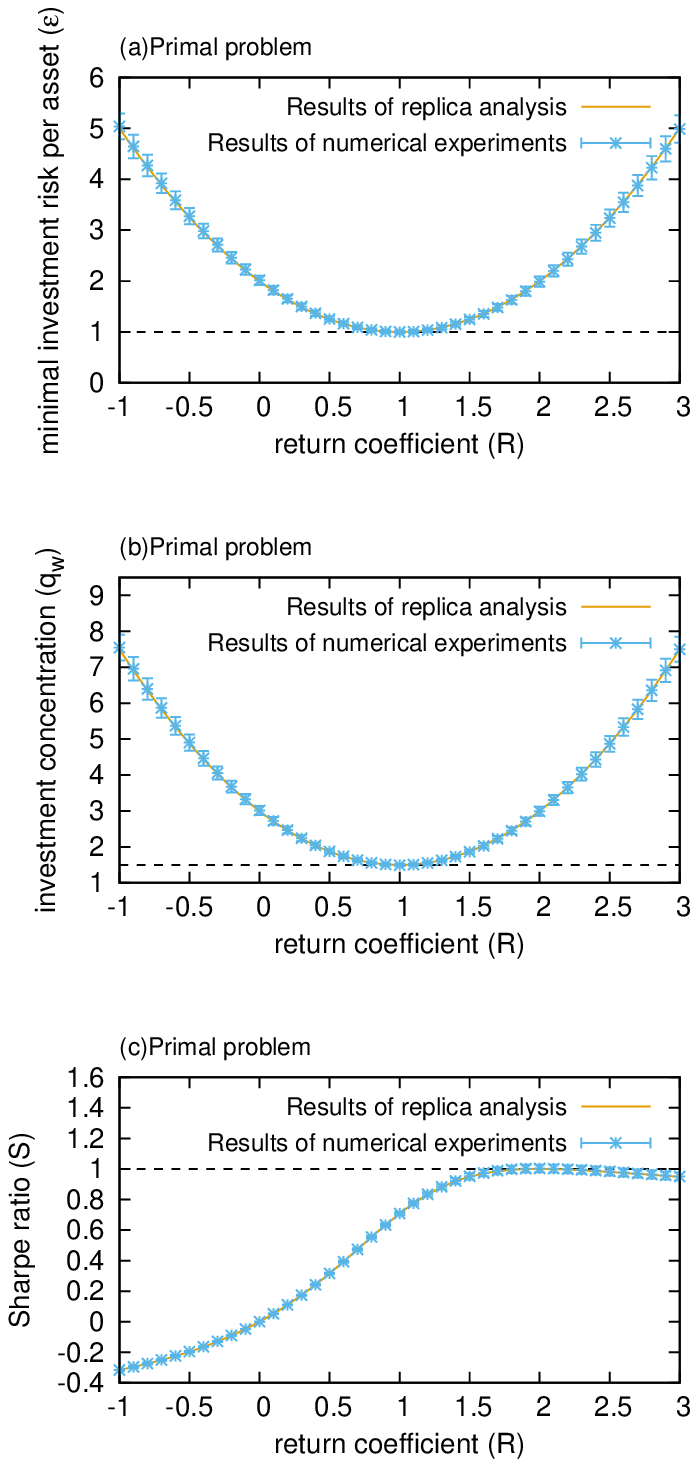}
\caption{
\label{Fig1}
Results of the replica analysis and the numerical experiments ($\a=p/N=3$).
The horizontal axis shows the return coefficient $R$, and the vertical axes show 
(a) the minimal investment risk per asset $\ve$, (b) the investment 
 concentration $q_w$, and (c) the Sharpe ratio $S$.
The solid (orange) lines indicate the results of the replica analysis for (a) 
 \sref{eq17}, (b) \sref{eq15}, and (c) \sref{eq20-1}. The (blue) asterisks with 
 the error bars indicate the results of the numerical simulation, and 
the dashed (black) lines indicate the results for (a) $\f{\sd^2(\a-1)}{2}$, (b) $\f{\a}{\a-1}$, and 
 (c) $S_{\max}$.
}
\end{center}
\end{figure}

\begin{figure}[tbh]
\begin{center}
\includegraphics[width=0.9\hsize,angle=0]{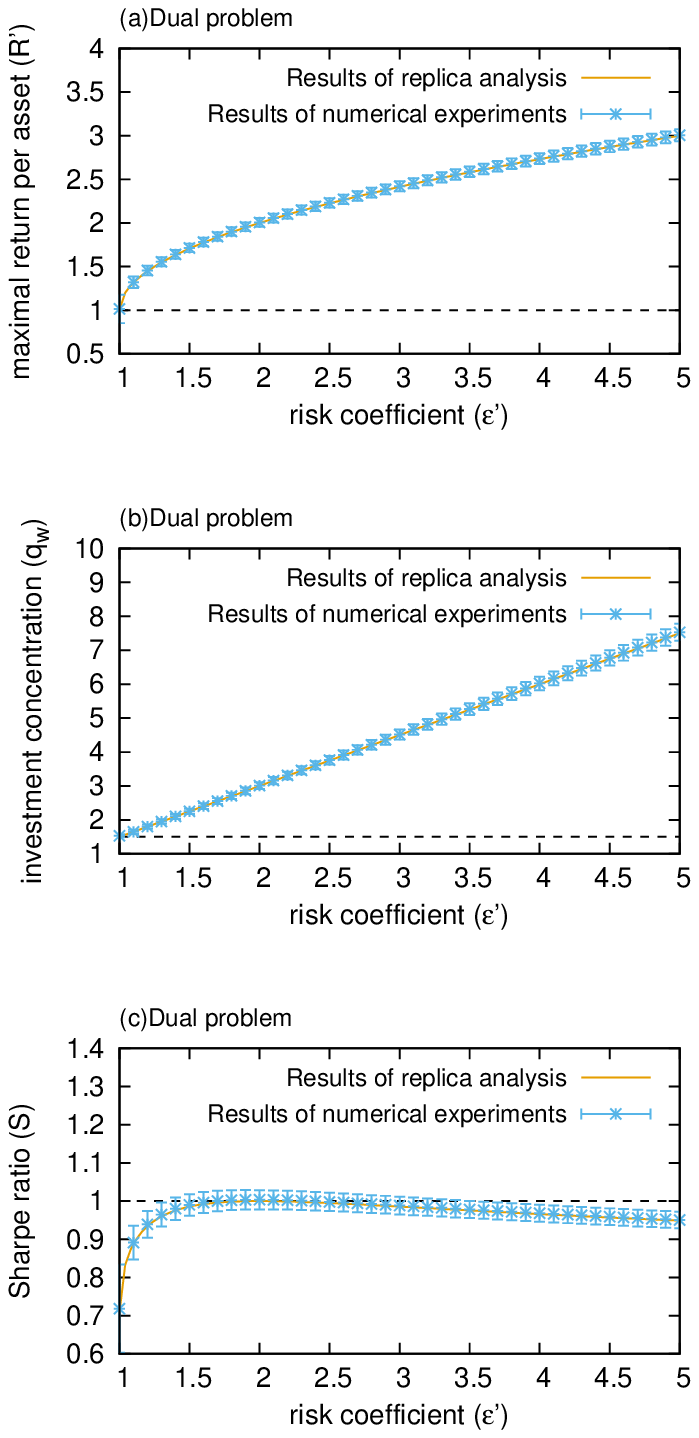}
\caption{
\label{Fig2}
Results obtained by the replica 
 analysis and the numerical experiments ($\a=p/N=3$).
The horizontal axis shows the risk coefficient $\ve'$, and the vertical axes show 
(a) the maximal return per asset $R'$, (b) the investment 
 concentration $q_w$, and (c) the Sharpe ratio $S$.
The solid (orange) lines show the results of the replica analysis for (a) \sref{eq30}, 
 (b) \sref{eq31}, and (c) \sref{eq34-1}. The (blue) asterisks with 
 the error bars indicate the results of the numerical simulation, and 
the dashed (black) lines indicate (a) $m$, (b) $\f{\a}{\a-1}$, and 
 (c) $S_{\max}$.
}
\end{center}
\end{figure}

Thus, in the numerical simulations, 
$N=1,000$ and $p=3,000$, i.e., $\a=p/N=3$,
the primal problem and the dual problem at $(\sd^2,m,\s^2)=(1,1,1)$ are 
examined. The sample size used in the estimation is $C=100$.
The results of the primal problem and the dual problem are shown in Figs. \ref{Fig1} and \ref{Fig2}, respectively.
In Fig. \ref{Fig1}, the horizontal axis 
shows the return coefficient $R$, and the vertical axes show (a) the minimal 
investment risk per asset $\ve$, (b) the investment concentration $q_w$, and 
(c) the Sharpe ratio $S$. Moreover, 
in Fig. \ref{Fig2},  
the horizontal axis shows the risk coefficient $\ve'$, and 
the vertical axes show (a) the maximal expected return per asset $R'$, 
(b) the investment concentration $q_w$, and 
(c) the Sharpe ratio $S$.
The solid (orange) lines indicate the results of the replica analysis, and 
the (blue) asterisks with the error bars show the numerical results. 
The dashed (black) lines in Fig. 
\ref{Fig1} indicate (a) $\f{\sd^2(\a-1)}{2}$, (b) $\f{\a}{\a-1}$, and (c) $S_{\max}$, and 
those in Fig. \ref{Fig2} indicate (a) $m$, (b) $\f{\a}{\a-1}$, and (c) $S_{\max}$.
As shown in these figures, the results derived by the proposed method are consistent with 
the numerical results, i.e., the effectiveness of the proposed approach is confirmed.


\section{Conclusion and future research\label{sec5}}

In the present paper, in order to extend the portfolio optimization problem of a quenched disordered system with only a budget constraint, 
{which} has was considered in our previous study \cite{Shinzato-SA2015}, we analyzed the portfolio optimization problem of a quenched disordered system with several constraints using replica analysis and discussed the primal-dual structure of the mean-variance model. In our previous studies \cite{Shinzato-qw-fixed2016,Shinzato-ve-fixed2016}, the primal-dual structure with respect to investment concentration and investment risk was assessed. In the present paper, the portfolio optimization problem minimizing the investment risk with budget and expected return constraints is regarded as the primal problem, and the portfolio optimization problem maximizing the expected return with budget and investment risk constraints is regarded as the dual problem. We clarified  the primal-dual structure in these two portfolio optimization problems. Similar to the annealed disordered system considered in general operations research studies, the minimal investment risk was confirmed to be a quadratic function with respect to the coefficient of the expected return constraint in the primal problem of a quenched disordered system. Moreover, in order to validate the effectiveness of the proposed method, we compared its results to those of numerical simulations and confirmed that there was good agreement.

In the future, 
since the randomness of the return rate 
in the study by Varga-Haszonits et al. \cite{VH} 
is different from that in the present paper, 
we need to consider the primal-dual problem in terms of 
the randomness used in that study and 
{theoretically develop a methodology for resolving the portfolio optimization problem}. 
Moreover, in such cases, we also need to verify the mathematical structure of the Sharpe ratio.

\appendix
\section{Lagrange multiplier method for the primal problem\label{app-a}}
In this appendix, we discuss the 
portfolio optimization problem by applying Lagrange's method of undetermined multipliers to the primal problem. First, 
the Lagrange undetermined multiplier function is given as follows:
\bea
L\eq\f{1}{2}\vec{w}^{\rm T}J\vec{w}+k\left(N-\vec{w}^{\rm T}\vec{e}\right)
+\theta\left(NR-\vec{w}^{\rm T}\vec{r}\right),
\eea
where the auxiliary variables $k$ and $\theta$ are used. Since $\pp{L}{\vec{w}}=0$, we obtain
\bea
\label{eq-a2}
\vec{w}\eq kJ^{-1}\vec{e}+\theta J^{-1}\vec{r},
\eea
and, since
$\pp{L}{k}=\pp{L}{\theta}=0$, we obtain
\bea
\label{eq-a3}
\left(
\begin{array}{c}
1\\
R
\end{array}
\right)
\eq
\left(
\begin{array}{cc}
\f{\vec{e}^{\rm T}J^{-1}\vec{e}}{N}&
\f{\vec{r}^{\rm T}J^{-1}\vec{e}}{N}\\
\f{\vec{r}^{\rm T}J^{-1}\vec{e}}{N}&
\f{\vec{r}^{\rm T}J^{-1}\vec{r}}{N}
\end{array}
\right)
\left(
\begin{array}{c}
k\\
\theta
\end{array}
\right),
\eea
where 
\bea
\left(
\begin{array}{c}
k\\
\theta
\end{array}
\right)
\eq\f{1}{D}
\left(
\begin{array}{cc}
\f{\vec{r}^{\rm T}J^{-1}\vec{r}}{N}&
-\f{\vec{r}^{\rm T}J^{-1}\vec{e}}{N}\\
-\f{\vec{r}^{\rm T}J^{-1}\vec{e}}{N}&
\f{\vec{e}^{\rm T}J^{-1}\vec{e}}{N}
\end{array}
\right)
\left(
\begin{array}{c}
1\\
R
\end{array}
\right),\\
\label{eq-a5}
D\eq
\left(\f{\vec{e}^{\rm T}J^{-1}\vec{e}}{N}\right)^2
\left\{
\f{\vec{r}^{\rm T}J^{-1}\vec{r}}{\vec{e}^{\rm T}J^{-1}\vec{e}}
-\left(
\f{\vec{r}^{\rm T}J^{-1}\vec{e}}{\vec{e}^{\rm T}J^{-1}\vec{e}}
\right)^2
\right\}.\qquad
\eea
In addition, from \sref{eq-a2}, we have
$J\vec{w}=k\vec{e}+\theta\vec{r}$. Then, 
from $\ve=\f{1}{2N}\vec{w}^{\rm T}J\vec{w}
=\f{k+R\theta}{2}$,
\bea
\ve\eq
\f{N}{2\vec{e}^{\rm T}J^{-1}\vec{e}}
\left\{1+
\f{\left(R-\f{\vec{r}^{\rm T}J^{-1}\vec{e}}{\vec{e}^{\rm T}J^{-1}\vec{e}}\right)^2}
{
\f{
\vec{r}^{\rm T}J^{-1}\vec{r}
}{\vec{e}^{\rm T}J^{-1}\vec{e}}
-
\left(\f{\vec{r}^{\rm T}J^{-1}\vec{e}}{\vec{e}^{\rm T}J^{-1}\vec{e}}\right)^2
}
\right\}
,
\eea
is solved. This finding depends on the given return rate matrix $X$ 
and mean vector $\vec{r}$. Next, we briefly consider the analysis 
of a quenched disordered system.

When $N$ is sufficiently large, both sides of 
\sref{eq-a3} are 
averaged  by {the mean vector $\vec{r}$.}   Then, we obtain
\bea
\left(
\begin{array}{c}
1\\
R
\end{array}
\right)
\eq
E_{\vec{r}}
\left[
\left(
\begin{array}{cc}
\f{\vec{e}^{\rm T}J^{-1}\vec{e}}{N}&
\f{\vec{r}^{\rm T}J^{-1}\vec{e}}{N}\\
\f{\vec{r}^{\rm T}J^{-1}\vec{e}}{N}&
\f{\vec{r}^{\rm T}J^{-1}\vec{r}}{N}
\end{array}
\right)
\left(
\begin{array}{c}
k\\
\theta
\end{array}
\right)
\right].
\eea
Each component of the coefficient matrix can be obtained independently as 
\bea
E_{\vec{r}}
\left[
\vec{r}^{\rm T}J^{-1}\vec{e}
\right]
\eq m\vec{e}^{\rm T}J^{-1}\vec{e},\\
E_{\vec{r}}
\left[
\vec{r}^{\rm T}J^{-1}\vec{r}
\right]
\eq m^2
\vec{e}^{\rm T}J^{-1}\vec{e}
+\s^2{\rm Tr}J^{-1}
\eea
in terms of the mean vector $\vec{r}$. Furthermore, using $N$ eigenvalues of the Wishart matrix 
$J=XX^{\rm T}\in{\bf R}^{N\times N}$, 
$\l_1,\cdots,\l_N$, 
since $\sum_{i=1}^N\l_i^{-1}=\vec{e}^{\rm T}J^{-1}\vec{e}={\rm Tr}J^{-1}$
holds, we have
\bea
\left(
\begin{array}{c}
1\\
R
\end{array}
\right)
\eq
\f{\vec{e}^{\rm T}J^{-1}\vec{e}}{N}
\left(
\begin{array}{cc}
1&m\\
m&m^2+\s^2
\end{array}
\right)
\left(
\begin{array}{c}
k\\
\theta
\end{array}
\right).
\eea
In addition, from \cite{Shinzato-SA2015}, we have
\bea
\label{eq-a11}
\f{1}{N}\vec{e}^{\rm T}J^{-1}\vec{e}\eq\f{1}{\sd^2(\a-1)},
\eea
i.e., we obtain
\bea
\ve\eq\f{\sd^2(\a-1)}{2}\left(1+\f{(R-m)^2}{\s^2}\right).
\eea
This finding is consistent with the 
results of the replica analysis in \sref{eq17}.
\section{Lagrange multiplier method for the dual problem\label{app-b}}
Here, we analyze the portfolio optimization by applying 
Lagrange's method of undetermined multipliers 
to the dual problem.
First, the Lagrange undetermined multiplier function is defined as follows:
\bea
L\eq \vec{r}^{\rm T}\vec{w}
+k(\vec{w}^{\rm T}\vec{e}-N)
+\theta
\left(N\ve'-\f{1}{2}\vec{w}^{\rm T}J\vec{w}\right),
\eea
where the auxiliary variables $k$ and $\theta$ are used. Since $\pp{L}{\vec{w}}=0$, we obtain
\bea
\label{eq-b2}
\vec{w}\eq \f{k}{\theta}J^{-1}\vec{e}+\f{1}{\theta}J^{-1}\vec{r},
\eea
and since $\pp{L}{k}=\pp{L}{\theta}=0$, we have
\bea
\label{eq-b3}
1\eq \f{k}{\theta}\f{\vec{e}^{\rm T}J^{-1}\vec{e}}{N}+\f{1}{\theta}\f{\vec{r}^{\rm T}J^{-1}\vec{e}}{N},\\
\ve'
\eq\f{1}{2}\left(
\f{k^2}{\theta^2}
\f{\vec{e}^{\rm T}J^{-1}\vec{e}}{N}+
2\f{k}{\theta^2}
\f{\vec{r}^{\rm T}J^{-1}\vec{e}}{N}
+\f{1}{\theta^2}
\f{\vec{r}^{\rm T}J^{-1}\vec{r}}{N}
\right),\nn
\eq\f{N}{2\vec{e}^{\rm T}J^{-1}\vec{e}}
\left[
\left(\f{k}{\theta}\f{\vec{e}^{\rm T}J^{-1}\vec{e}}{N}+\f{1}{\theta}\f{\vec{r}^{\rm T}J^{-1}\vec{e}}{N}\right)^2
\right.\nn
&&\left.
+\f{1}{\theta^2}
\left(\f{\vec{e}^{\rm T}J^{-1}\vec{e}}{N}\right)^2
\left\{
\f{\vec{r}^{\rm T}J^{-1}\vec{r}}{\vec{e}^{\rm T}J^{-1}\vec{e}}
-\left(
\f{\vec{r}^{\rm T}J^{-1}\vec{e}}{\vec{e}^{\rm T}J^{-1}\vec{e}}
\right)^2
\right\}
\right].\qquad
\label{eq-b4}
\eea
In summary, 
\bea
\theta
\eq\f{\sqrt{D}}
{\sqrt{
2\ve'
\f{\vec{e}^{\rm T}J^{-1}\vec{e}}{N}-1
}},
\eea
is obtained, where $D$ is as defined in \sref{eq-a5}. Moreover, 
\bea
k\eq
\f{N}{\vec{e}^{\rm T}J^{-1}\vec{e}}
\left(
\theta-
\f{\vec{r}^{\rm T}J^{-1}\vec{e}}{N}
\right)
\eea
is derived from 
\sref{eq-b3}. In addition, using \sref{eq-b2}, 
from $N\ve'=\f{1}{2}\vec{w}^{\rm 
T}\left(\f{k}{\theta}\vec{e}+\f{1}{\theta}\vec{r}\right)=\f{N}{2\theta}(k+R')$, we have
\bea
R'\eq{2\ve'\theta-k}\nn
\eq
\sqrt{\f{\vec{r}^{\rm T}J^{-1}\vec{r}}
{\vec{e}^{\rm T}J^{-1}\vec{e}}
-
\left(\f{\vec{r}^{\rm T}J^{-1}\vec{e}}
{\vec{e}^{\rm T}J^{-1}\vec{e}}\right)^2
}
\sqrt{
\f{2\ve'\vec{e}^{\rm T}J^{-1}\vec{e}}{N}-1
}\nn
&&
+
\f{\vec{r}^{\rm T}J^{-1}\vec{e}}
{\vec{e}^{\rm T}J^{-1}\vec{e}}.
\eea
This finding depends on a given return rate matrix $X$ and mean vector $\vec{r}$. Next, 
we briefly consider the analysis of a quenched disordered system.

If $N$ is sufficiently large, both sides in Eqs. (\ref{eq-b3}) and 
(\ref{eq-b4}) are averaged over the mean vector. Then, we obtain 
\bea
1\eq\f{\vec{e}^{\rm T}J^{-1}\vec{e}}{N}
\f{k+m}{\theta},\\
\ve'\eq
\f{\vec{e}^{\rm T}J^{-1}\vec{e}}{2N}
\left(\f{k^2}{\theta^2}+2m\f{k}{\theta^2}+\f{1}{\theta^2}(m^2+\s^2)\right).
\eea
In other words, 
\bea
k\eq
\f{N\theta}{\vec{e}^{\rm T}J^{-1}\vec{e}}-m,\\
\theta\eq\f{\s\vec{e}^{\rm T}J^{-1}\vec{e}}{N}
\f{1}{\sqrt{
\f{2\ve'\vec{e}^{\rm T}J^{-1}\vec{e}}{N}
-1}},
\eea
are derived.  Thus, $R'={2\ve'\theta-k}$ is solved as follows:
\bea
R'\eq
\left(
2\ve'-
\f{N}{\vec{e}^{\rm T}J^{-1}\vec{e}}
\right)\theta
+m\nn
\eq m+\s\sqrt{
\f{2\ve'\vec{e}^{\rm T}J^{-1}\vec{e}}{N}
-1},
\eea
and using \sref{eq-a11}, we can calculate 
$\f{1}{N}\vec{e}^{\rm T}J^{-1}\vec{e}=\f{1}{\sd^2(\a-1)}$,
\bea
R'\eq m+\s\sqrt{\f{2\ve'}{\sd^2(\a-1)}-1}.
\eea
This finding is consistent with the results derived by 
the replica analysis in \sref{eq30}.
\section{Replica approach for the primal problem\label{app-c}}
In this appendix, we demonstrate a replica analysis of a quenched 
disordered system of the primal problem. 
Following previous studies, 
when $n\in{\bf Z}$, the configuration average of the $n$th power of the 
partition function $Z(R,X,\vec{r})$, $E_{X,\vec{r}}
\left[Z^n(R,X,\vec{r})\right]$ is expanded as follows:
\bea
&&
E_{X,\vec{r}}
\left[Z^n(R,X,\vec{r})\right]\nn
\eq\f{1}{(2\pi)^{\f{Nn}{2}+pn}}
\mathop{\rm Extr}_{\vec{k},\vec{\theta}}
\area \prod_{a=1}^nd\vec{w}_ad\vec{u}_ad\vec{v}_a\nn
&&
E_{X,\vec{r}}
\left[
\exp
\left(
-\f{\b}{2}\sum_{a=1}^n\sum_{\mu=1}^p
v_{\mu a}^2
+i\sum_{a=1}^n
\sum_{\mu=1}^pu_{\mu a}v_{\mu a}
\right.
\right.\nn
&&-\f{i}{\sqrt{N}}
\sum_{i=1}^N
\sum_{\mu=1}^px_{i\mu}
\sum_{a=1}^nu_{\mu a}w_{ia}
+\sum_{a=1}^nk_a
\left(\sum_{i=1}^Nw_{ia}-N\right)\nn
&&\left.
\left.
+\sum_{a=1}^n\theta_a
\left(\sum_{i=1}^Nr_iw_{ia}-NR\right)
\right)
\right]\nn
\eq\f{1}{(2\pi)^{\f{Nn}{2}+pn}}
\mathop{\rm Extr}_{\vec{k},\vec{\theta},Q_w,\tilde{Q}_w}
\area \prod_{a=1}^nd\vec{w}_ad\vec{u}_ad\vec{v}_a\nn
&&\exp
\left(
-\f{\b}{2}
\sum_{\mu=1}^p\sum_{a=1}^nv_{\mu a}^2+
i\sum_{\mu=1}^p
\sum_{a=1}^n
u_{\mu a}v_{\mu a}
\right.\nn
&&
-\f{\sd^2}{2}\sum_{\mu=1}^p\sum_{a=1}^n\sum_{b=1}^nq_{wab}u_{\mu a}u_{\mu b}
-N\sum_{a=1}^nk_a-NR\sum_{a=1}^n\theta_a\nn
&&+Nm\sum_{a=1}^n\theta_a+\f{N\s^2}{2}
\sum_{a=1}^n\sum_{b=1}^n\theta_a\theta_bq_{wab}
+\sum_{i=1}^N\sum_{a=1}^nk_aw_{ia}
\nn
&&
\left.
-\f{1}{2}\sum_{a=1}^n\sum_{b=1}^n\tilde{q}_{wab}
\left(\sum_{i=1}^Nw_{ia}w_{ib}-q_{wab}\right)
\right),
\eea
where 
$\vec{w}_a=(w_{1a},\cdots,w_{Na})^{\rm T}\in{\bf 
R}^N,\vec{u}_a=(u_{1a},\cdots,u_{pa})^{\rm T}\in{\bf R}^p,
\vec{v}_a=(v_{1a},\cdots,v_{pa})^{\rm T}\in{\bf R}^p,(a=1,\cdots,n)$, 
and 
\bea
E_X\left[
e^{-\f{ix_{i\mu}A}{\sqrt{N}}}
\right]
&\simeq& e^{-\f{\sd^2}{2N}A^2},\\
E_{\vec{r}}\left[
e^{r_iB}
\right]&\simeq& e^{mB+\f{\s^2}{2}B^2},
\eea
are used as the configuration average on $x_{i\mu},r_i$. Moreover, $\vec{k}=(k_1,\cdots,k_n)^{\rm T}\in{\bf R}^n,
\vec{\theta}=(\theta_1,\cdots,\theta_n)^{\rm T}\in{\bf R}^n,
Q_w=\left\{q_{wab}\right\}\in{\bf R}^{n\times n},
\tilde{Q}_w=\left\{\tilde{q}_{wab}\right\}\in{\bf R}^{n\times n}
$ are used. Then, as the number of assets $N$ approaches infinity, we have
\bea
&&\lim_{N\to\infty}\f{1}{N}\log E_{X,\vec{r}}
\left[Z^n(R,X,\vec{r})\right]\nn
\eq
\mathop{\rm Extr}_{\vec{k},\vec{\theta},Q_w,\tilde{Q}_w}
\left\{
\f{1}{2}{\rm Tr}Q_w\tilde{Q}_w-\vec{k}^{\rm 
T}\vec{e}-(R-m)\vec{\theta}^{\rm T}\vec{e}
\right.\nn
&&
+\f{\s^2}{2}\vec{\theta}^{\rm T}Q_w
\vec{\theta}
-\f{\a}{2}\log\det\left|
I+\b\sd^2 Q_w
\right|\nn
&&
\left.-\f{1}{2}\log\det\left|\tilde{Q}_w\right|
+\f{1}{2}\vec{k}^{\rm T}\tilde{Q}_w^{-1}\vec{k}
\right\},
\eea
where $\a=p/N\sim O(1)$, and 
the identity matrix $I\in{\bf R}^{n\times n}$ and unit vector 
$\vec{e}=(1,\cdots,1)^{\rm T}\in{\bf R}^n$
are used. Then, based on the ansatz of the replica symmetry solution, 
with respect to $a,b=1,2,\cdots,n$,
\bea
q_{wab}\eq
\left\{
\begin{array}{ll}
\chi_w+q_w&a=b\\
q_w&a\ne b
\end{array}
\right.,\\
\tilde{q}_{wab}\eq
\left\{
\begin{array}{ll}
\tilde{\chi}_w-\tilde{q}_w&a=b\\
-\tilde{q}_w&a\ne b
\end{array}
\right.,\\
k_a\eq k,\\
\theta_a\eq \theta,
\eea
are set, and 
\bea
\phi(R)\eq\lim_{n\to0}\pp{}{n}
\left\{
\lim_{N\to\infty}
\f{1}{N}
\log E_{X,\vec{r}}
\left[Z^n(R,X,\vec{r})\right]
\right\}
\nn
\eq\mathop{\rm Extr}_\Theta
\left\{
\f{1}{2}(\chi_w+q_w)(\tilde{\chi}_w-\tilde{q}_w)+\f{q_w\tilde{q}_w}{2}\right.\nn
&&
-k-(R-m)\theta
+\f{\s^2\theta^2}{2}\chi_w
-\f{\a}{2}\log(1+\b\sd^2\chi_w)\nn
&&\left.-\f{\a\b\sd^2 q_w}{2(1+\b\sd^2\chi_w)}
-\f{1}{2}\log\tilde{\chi}_w+\f{\tilde{q}_w+k^2}{2\tilde{\chi}_w}
\right\},
\eea
is evaluated. Then, as the extremal conditions, we obtain
\bea
\pp{\phi(R)}{k}\eq-1+\f{k}{\tilde{\chi}_w}=0,\\
\pp{\phi(R)}{\theta}\eq-(R-m)+\s^2\chi_w\theta=0,\\
\pp{\phi(R)}{\chi_w}\eq\f{1}{2}(\tilde{\chi}_w-\tilde{q}_w)+\f{\s^2\theta^2}{2}
-\f{\a\b\sd^2}{2(1+\b\sd^2\chi_w)}\nn
&&+
\f{\a\b^2\sd^4q_w}{2(1+\b\sd^2\chi_w)^2}=0,\\
\pp{\phi(R)}{q_w}\eq\f{1}{2}(\tilde{\chi}_w-\tilde{q}_w)+\f{\tilde{q}_w}{2}
-\f{\a\b\sd^2}{2(1+\b\sd^2\chi_w)}=0,\qquad\\
\pp{\phi(R)}{\tilde{\chi}_w}\eq\f{1}{2}({\chi}_w+{q}_w)-\f{1}{2\tilde{\chi}_w}-
\f{\tilde{q}_w+k^2}{2\tilde{\chi}_w^2}=0,\\
\pp{\phi(R)}{\tilde{q}_w}\eq-\f{1}{2}({\chi}_w+{q}_w)+\f{q_w}{2}+\f{1}{2\tilde{\chi}_w}=0,
\eea
and 
\bea
\chi_w\eq\f{1}{\b\sd^2(\a-1)},\\
q_w\eq\f{\a}{\a-1}\left(1+\f{(R-m)^2}{\s^2}\right).
\eea
Furthermore, the minimal investment risk per asset $\ve$ is
\bea
\ve\eq-\lim_{\b\to\infty}\pp{\phi(R)}{\b}\nn
\eq
\lim_{\b\to\infty}
\left\{
\f{\a\sd^2\chi_w}{2(1+\b\sd^2\chi_w)}
+\f{\a\sd^2q_w}{2(1+\b\sd^2\chi_w)^2}
\right\}\nn
\eq\f{\sd^2(\a-1)}{2}
\left(1+\f{(R-m)^2}{\s^2}\right).
\eea

\section{Replica approach for the dual problem\label{app-d}}
In this appendix, 
we explain in detail the replica analysis of a quenched disordered system involving the dual problem. Following the discussion in the 
above appendix, when $n\in{\bf Z}$, 
the configuration average of $n$th power of the 
partition function 
$Z(\ve',X,\vec{r})$, 
$E_{X,\vec{r}}
\left[
Z^n(\ve',X\vec{r})
\right]
$ is expanded as follows: 
\bea
&&E_{X,\vec{r}}
\left[
Z^n(\ve',X\vec{r})
\right]\nn
\eq\f{1}{(2\pi)^{\f{Nn}{2}+pn}}
\mathop{\rm Extr}_{\vec{k},\vec{\theta}}
\area\prod_{a=1}^nd\vec{w}_ad\vec{u}_ad\vec{v}_a\nn
&&E_{X,\vec{r}}
\left[
\exp\left(
\b\sum_{a=1}^n
\sum_{i=1}^Nr_iw_{ia}+i\sum_{a=1}^n\sum_{\mu=1}^pu_{\mu a}v_{\mu a}
\right.
\right.\nn
&&-\f{i}{\sqrt{N}}\sum_{i=1}^N\sum_{\mu=1}^p
x_{i\mu}\sum_{a=1}^nu_{\mu a}w_{ia}+
\sum_{a=1}^nk_a
\left(\sum_{i=1}^Nw_{ia}-N\right)\nn
&&\left.\left.
+\sum_{a=1}^n\theta_a
\left(
N\ve'-\f{1}{2}
\sum_{\mu=1}^pv_{\mu a}^2
\right)
\right)\right].
\eea
As in the previous discussion, 
as the number of assets $N$ approaches infinity, we obtain
\bea
&&
\lim_{N\to\infty}\f{1}{N}\log E_{X,\vec{r}}
\left[Z^n(\ve',X,\vec{r})\right]\nn
\eq\mathop{\rm Extr}_{\vec{k},\vec{\theta},Q_w,\tilde{Q}_w}
\left\{
\f{1}{2}{\rm Tr}Q_w\tilde{Q}_w
-\vec{k}^{\rm T}\vec{e}+\ve'\vec{\theta}^{\rm T}\vec{e}
+n\b m
\right.\nn
&&+\f{\s^2\b^2}{2}\vec{e}^{\rm 
T}Q_w\vec{e}-\f{\a}{2}\log\det\left|I+\sd^2\Theta Q_w\right|\nn
&&\left.
-\f{1}{2}\log\det\left|\tilde{Q}_w\right|
+\f{1}{2}\vec{k}^{\rm T}\tilde{Q}_w^{-1}\vec{k}
\right\},
\eea
where $\Theta={\rm diag}(\theta_1,\cdots,\theta_n)\in{\bf 
R}^{n\times n}$ is used. Then, based on the assumption of a replica 
symmetry solution, we obtain
\bea
\phi(\ve')\eq\lim_{n\to0}\pp{}{n}
\left\{
\lim_{N\to\infty}\f{1}{N}\log E_{X,\vec{r}}
\left[Z^n(\ve',X,\vec{r})\right]
\right\}\nn
\eq
\mathop{\rm Extr}_\Theta
\left\{
\f{1}{2}(\chi_w+q_w)(\tilde{\chi}_w-\tilde{q}_w)+\f{q_w\tilde{q}_w}{2}
\right.\nn
&&-k+\theta\ve'+
\b m+\f{\s^2\b^2}{2}\chi_w
-\f{\a}{2}\log(1+\theta\sd^2\chi_w)
\nn
&&
\left.
-\f{\a\theta\sd^2q_w}{2(1+\theta\sd^2\chi_w)}
-\f{1}{2}\log\tilde{\chi}_w
+\f{\tilde{q}_w+k^2}{2\tilde{\chi}_w}
\right\}.
\eea
Then, as the extremal conditions, we obtain
\bea
\pp{\phi(\ve')}{k}\eq-1+\f{k}{\tilde{\chi}_w}=0,\\
\pp{\phi(\ve')}{\theta}\eq\ve'-\f{\a\sd^2\chi_w}{2(1+\theta\sd^2\chi_w)}
-\f{\a\sd^2q_w}{2(1+\theta\sd^2\chi_w)}\nn
&&
+\f{\a\theta\sd^4q_w\chi_w}{2(1+\theta\sd^2\chi_w)^2}=0,\\
\pp{\phi(\ve')}{\chi_w}\eq
\f{1}{2}(\tilde{\chi}_w-\tilde{q}_w)+\f{\s^2\b^2}{2}-\f{\a\theta\sd^2}{2(1+\theta\sd^2\chi_w)}\nn
&&
+\f{\a\theta^2\sd^4q_w}{2(1+\theta\sd^2\chi_w)^2}=0,\\
\pp{\phi(\ve')}{q_w}\eq
\f{1}{2}(\tilde{\chi}_w-\tilde{q}_w)+\f{\tilde{q}_w}{2}-\f{\a\theta\sd^2}{2(1+\theta\sd^2\chi_w)}=0,\qquad\\
\pp{\phi(\ve')}{\tilde{\chi}_w}\eq\f{1}{2}(\chi_w+q_w)-\f{1}{2\tilde{\chi}_w}-\f{\tilde{q}_w+k^2}{2\tilde{\chi}_w^2}=0,\\
\pp{\phi(\ve')}{\tilde{q}_w}\eq-\f{1}{2}(\chi_w+q_w)+\f{q_w}{2}+\f{1}{2\tilde{\chi}_w}=0,
\eea
and 
\bea
\chi_w\eq\f{1}{\theta\sd^2(\a-1)},\\
q_w\eq\f{\a}{\a-1}\left(1+\b^2\s^2\chi_w^2\right).
\eea
Substituting the above into $\ve'=\f{\a\sd^2\chi_w}{2(1+\theta\sd^2\chi_w)}+
\f{\a\sd^2q_w}{2(1+\theta\sd^2\chi_w)^2}$, we then obtain
\bea
\label{eq-d12}
\ve'\eq\f{1}{2\theta}+\f{\sd^2(\a-1)}{2}\left(1+\b^2\s^2\chi_w^2\right),
\eea
and 
\bea
\b\s\chi_w\eq\sqrt{\f{2}{\sd^2(\a-1)}
\left(\ve'-\f{1}{2\theta}\right)-1
},
\eea
are evaluated. Thus, the maximal expected return per asset $R'$ is 
calculated as follows:
\bea
R'\eq\lim_{\b\to\infty}
\pp{\phi(\ve')}{\b}\nn
\eq m+\s\lim_{\b\to\infty}\b\s\chi_w\nn
\eq m+\s\sqrt{\f{2\ve'}{\sd^2(\a-1)}
-1
},
\eea
where, from $\b\s\chi_w\sim O(1)$, 
$\b/\theta\sim O(1)$ is used. In addition, 
from 
\sref{eq-d12},
\bea
\b\s\chi_w\eq-\sqrt{\f{2}{\sd^2(\a-1)}
\left(\ve'-\f{1}{2\theta}\right)-1
},
\eea
is also derived, as $\b\to-\infty$,
the minimal expected return per asset under the investment risk is fixed, 
and $R''$ is obtained as follows:
\bea
R''\eq m-\s\sqrt{\f{2\ve'}{\sd^2(\a-1)}
-1
}.
\eea

\section{Annealed disordered approach for the primal problem\label{app-e}}
Here, the typical behavior of an annealed disordered system involving the primal problem 
is discussed.
First, following the analytical procedure of operations research, 
the expected investment risk $E_X
\left[
{\cal H}(\vec{w}|X)
\right]$ is calculated as follows: 
\bea
\label{eq-e1}
E_X
\left[
{\cal H}(\vec{w}|X)
\right]
\eq\f{1}{2}\vec{w}^{\rm T}
E_X
\left[
XX^{\rm T}
\right]\vec{w}\nn
\eq\f{\sd^2\a}{2}\sum_{i=1}^Nw_{i}^2.
\eea
Then, the object function of Lagrange's method of undetermined 
multipliers is prepared as follows:
\bea
L^{\rm OR}\eq\f{\sd^2\a}{2}\sum_{i=1}^Nw_{i}^2+k^{\rm OR}
\left(N-\vec{w}^{\rm T}\vec{e}\right)\nn
&&+\theta^{\rm OR}
\left(NR-\vec{w}^{\rm T}\vec{r}\right),
\eea
where the auxiliary variables $k^{\rm OR},\theta^{\rm OR}$ are used. Thus, 
from $\pp{L^{\rm OR}}{w_i}=0$, we obtain
\bea
w_i\eq\f{k^{\rm OR}+\theta^{\rm OR} r_i}{\sd^2\a}.
\eea
Furthermore, from 
$\pp{L^{\rm OR}}{k^{\rm OR}}=\pp{L^{\rm OR}}{\theta^{\rm OR}}=0$,
\bea
\left(
\begin{array}{c}
k^{\rm OR}\\
\theta^{\rm OR}
\end{array}
\right)
\eq
{\sd^2\a}
\left(
\begin{array}{cc}
1&\f{1}{N}\sum_{i=1}^Nr_i\\
\f{1}{N}\sum_{i=1}^Nr_i&
\f{1}{N}\sum_{i=1}^Nr_i^2
\end{array}
\right)^{-1}
\left(
\begin{array}{c}
1\\
R
\end{array}
\right)\nn
\eq
\f{\sd^2\a}{m^2+\s^2-m^2}
\left(
\begin{array}{cc}
m^2+\s^2&-m\\
-m&1
\end{array}
\right)
\left(
\begin{array}{c}
1\\
R
\end{array}
\right),\qquad
\eea
is assessed, where when $N$ is sufficiently large, we have 
\bea
\f{1}{N}\sum_{i=1}^Nr_i\eq m,\\
\f{1}{N}\sum_{i=1}^Nr_i^2\eq m^2+\s^2.
\eea
Then, the typical behavior of an annealed disordered system is 
obtained as follows:
\bea
\ve^{\rm OR}
\eq\f{\sd^2\a}{2}\f{1}{N}
\sum_{i=1}^N
\left(\f{k^{\rm OR}+\theta^{\rm OR}r_i}{\sd^2\a}\right)^2\nn
\eq\f{1}{2\sd^2\a}\left((k^{\rm OR})^2+2k^{\rm OR}\theta^{\rm OR}m
+
(\theta^{\rm OR})^2
(m^2+\s^2)
\right)\nn
\eq\f{\sd^2\a}{2}\left(1+\f{(R-m)^2}{\s^2}\right).
\eea

\section{Annealed disordered approach for the dual problem\label{app-f}}
Here, the typical behavior of an annealed disordered system involving the dual problem is also 
discussed. First, followed by the analytical procedure of operations research, the object function of 
Lagrange's method of undetermined multipliers is
defined as follows:
\bea
L^{\rm OR}
\eq\vec{r}^{\rm T}\vec{w}
+k^{\rm OR}
\left(\vec{w}^{\rm T}\vec{e}-N\right)\nn
&&+\theta^{\rm OR}
\left(N\ve'-\f{\sd^2\a}{2}\sum_{i=1}^Nw_i^2\right),
\eea
where, as in the primal problem discussed above, the 
expected investment risk in \sref{eq-e1}, $\f{\sd^2\a}{2}\sum_{i=1}^Nw_i^2$ is used.
Thus, from $\pp{L^{\rm OR}}{w_i}=0$, we obtain
\bea
w_i\eq\f{k^{\rm OR}+r_i}{\sd^2\a \theta^{\rm OR}}.
\eea
Moreover, from $\pp{L^{\rm OR}}{k^{\rm OR}}=\pp{L^{\rm OR}}{\theta^{\rm OR}}=0$, we obtain
\bea
1\eq\f{k^{\rm OR}+m}{\sd^2\a\theta^{\rm OR}},\\
\ve'\eq\f{\sd^2\a}{2\sd^4\a^2(\theta^{\rm OR})^2}
((k^{\rm OR})^2+2
k^{\rm OR}m+m^2+\s^2
)\nn
\eq\f{\sd^2\a}{2}\left(1+
\left(
\f{\s}{\sd^2\a\theta^{\rm OR}}
\right)^2
\right).\label{eq-f4}
\eea
Based on this and $R'=\f{1}{N}\sum_{i=1}^Nr_i
\left(
\f{k^{\rm 
OR}+r_i}{\sd^2\a\theta^{\rm OR}}\right)=\f{k^{\rm 
OR}m+m^2+\s^2}{\sd^2\a\theta^{\rm OR}}$, we obtain 
\bea
R'\eq m+\s\f{\s}{\sd^2\a\theta^{\rm OR}}\nn
\eq m+\s\sqrt{\f{2\ve'}{\sd^2\a}-1},
\eea
where, from \sref{eq-f4}, we use
\bea
\f{\s}{\sd^2\a\theta^{\rm OR}}
\eq
\sqrt{\f{2\ve'}{\sd^2\a}-1}.
\eea

\end{document}